\documentclass[a4paper,12pt,UKenglish]{article}

\usepackage[utf8]{inputenc}
\usepackage{natbib,babel,ifpdf,fullpage,bm,graphicx,ifthen,amsmath,lmodern,url}

\newcommand{\Sidak}{\v Sid\'ak }
\newcommand{\al}{\alpha_{\text{loc}}}
\newcommand{\Meff}{M_{\text{eff}}}
\newcommand\cov{\operatorname{Cov}}
\newcommand\tr{\operatorname{tr}}
\newcommand\var{\operatorname{Var}}
 
\newcommand{\xg}[1][]{\ifthenelse{\equal{#1}{}}{X_\text{g}}{\bm X_{\text{g}#1}}}
\newcommand{\xe}[1][]{\ifthenelse{\equal{#1}{}}{X_\text{e}}{\bm X_{\text{e}#1}}}
\newcommand{\x}[1][]{\ifthenelse{\equal{#1}{}}{X}{\bm X_{#1}}}
\newcommand{\y}[1][]{\ifthenelse{\equal{#1}{}}{\bm Y}{Y_{#1}}}
\newcommand{\bg}[1][]{\ifthenelse{\equal{#1}{}}{\bm\beta_\text{g}}{\beta_{\text{g}#1}}}
\newcommand{\be}[1][]{\ifthenelse{\equal{#1}{}}{\bm\beta_\text{e}}{\beta_{\text{e}#1}}}

\ifpdf 
   \DeclareGraphicsExtensions{.pdf,.eps,.png} 
\fi

\begin{document}
\title{Efficient and powerful familywise error control in genome-wide association studies using generalized linear models}

\author{K. K. Halle$^{1,2}$, {\O}. Bakke$^1$,  S. Djurovic$^{3,4}$, A. Bye$^{7}$, E. Ryeng$^8$,\\ U. Wisl{\o}ff$^7$, O. A. Andreassen$^{5,6}$ and M. Langaas$^{1}$\\ \\
\footnotesize\parbox{.88\linewidth}{$^1$Department of Mathematical Sciences, Norwegian University of Science
and Technology, NO-7491 Norway.
$^2$Liaison Committee between the Central Norway Regional Health Authority (RHA) and the Norwegian University of Science and Technology (NTNU), Trondheim, Norway.
$^3$NORMENT, K. G. Jebsen Centre for Psychosis Research, Department of Clinical Science, University of Bergen, Bergen, Norway.
$^4$Department of Medical Genetics, Oslo University Hospital, Oslo, Norway.
$^5$NORMENT, K. G. Jebsen Centre for Psychosis Research, Division of Mental Health and Addiction, Oslo University Hospital and Institute of Clinical Medicine, University of Oslo, Oslo, Norway.
$^6$Institute of Clinical Medicine,  University of Oslo, NO-0318 Oslo, Norway.
$^7$K. G. Jebsen Centre for Exercise in Medicine, Department of Circulation and Medical Imaging, Faculty of Medicine, Norwegian University of Science and Technology, Trondheim, Norway.
$^8$Department of Cancer Research and Molecular Medicine, Norwegian University of Science and Technology, NO-7491 Trondheim, Norway.}
}
\date{21 December 2016}
\maketitle

\begin{abstract}
In genetic association studies, detecting phenotype--genotype association is a primary goal. We assume that the relationship between the data -- phenotype, genetic markers and environmental covariates -- can be modelled by a generalized linear model (GLM). The inclusion of environmental covariates makes it possible to account for important confounding factors, such as sex and population substructure. A multivariate score statistic, which under the complete null hypothesis of no phenotype--genotype association asymptotically has a multivariate normal distribution with a covariance matrix that can be estimated from the data, is used to test a large number of genetic markers for association with the phenotype. We stress the importance of controlling the familywise error rate (FWER), and use the asymptotic distribution of the multivariate score test statistic to find a local significance level for the individual test. Using real data (from one study on schizophrenia and bipolar disorder and one on maximal oxygen uptake) and constructed correlated structures, we show that our method is a powerful alternative to the popular Bonferroni and \Sidak methods. For GLMs without environmental covariates, we show that our method is an efficient alternative to permutation methods for multiple testing. Further, we show that if environmental covariates and genetic markers are uncorrelated, the estimated covariance matrix of the score test statistic can be approximated by the estimated correlation matrix for just the genetic markers. As byproducts of our method, an effective number of independent tests can be defined, and FWER-adjusted $p$-values can be calculated as an alternative to using a local significance level.
\end{abstract}


\section{Introduction}\label{sec:intro}

In genome-wide association (GWA) studies the aim is to test for association between genetic markers and a phenotype. A large number of markers are tested, and it is important to control the overall Type I error rate. Our focus is on controlling the familywise error rate (FWER). Multiple testing correction methods may achieve this goal by estimating a local significance level for the individual tests. In this work we present a new method, {\em the order $k$ FWER-approximation method}, for finding a local significance level in multiple hypothesis testing for correlated common variants, as is often observed in GWA studies.

Assume that we have collected independent individual observations in a case--control, cohort or cross-sectional study. The phenotype of interest can be continuous or discrete. We consider biallelic genetic markers, giving three possible genotypes. For each genetic marker we specify a hypothesis situation, where the null hypothesis is of the type ``no association between the phenotype and genetic marker'' and we have a two sided alternative. We will model the data using a generalized linear regression model (GLM) with phenotype as response (outcome), genotype as the independent variable of interest (exposure), and possibly non-genetical, referred to as environmental, independent covariates (not of interest) in the model. In epidemiological studies, a confounder is a common factor which is associated with both the exposure and outcome. In GWA studies, population substructure may be associated with both the exposure (genotype) and outcome (phenotype) and therefore may be a confounding factor and need to be adjusted for in the analysis. Population stratification can be adjusted for by including principal components of the genotype covariance matrix of the individuals as covariates in the model \citep{Price2006}. As test statistics for the multiple hypothesis problem we use the score test statistics to evaluate the genotype contribution to the model for each genetic marker separately. It is known that the vector of separate score test statistics asymptotically follows a multivariate normal distribution with a covariance matrix that can be estimated using key features of the fitted GLM model and the genetic markers \citep{Schaid2002, SeamanMyhsok2005}. This has also been a key ingredient in the work of \cite{ConneelyBoehnke2007}.

Further, we show that for the special case when no environmental covariates are present or when environmental and genetic covariates are observed to be independent, the estimated correlation matrix between score test statistics can be approximated by the estimated correlation matrix between the genetic markers.

In a multiple testing situation with $m$ tests the familywise error rate can be controlled at level $\alpha$ by specifying a local $p$-value cut-off, $\al$, to be used for all the $m$ hypothesis tests. Inspired from the work of \cite{MoskvinaSchmidt2008} and \cite{DickhausStange2013} we will use an approximation to the $m$-dimensional asymptotic simultaneous multivariate normal distribution of the score test statistics vector to estimate $\al$. The $\al$ estimate can be used to define an effective number of independent tests, and our FWER-approximation can be used to compute FWER-adjusted $p$-values.

The order $k$ FWER-approximation method is more powerful than the \Sidak method (which assumes that the score test statistics are independent across markers) and the Bonferroni method (which is valid for all dependence structures between the score test statistics). Further, it is more efficient and more widely applicable than the method of \cite{ConneelyBoehnke2007}. In Section \ref{sec:discuss} we will see that the method of \cite{ConneelyBoehnke2007}  is built on numerical integration in $m$ dimensions and is computationally intensive. 

The Westfall--Young permutation procedure is known to have asymptotically optimal power for a broad class of problems, including block-dependent and sparse dependence structure \citep{MeinshausenMaathuisBuhlman2011}. However, this method is computer intensive and to have a valid permutation test, the assumption of exchangeability needs to be satisfied \citep{Commenges2003}. This assumption is in general not satisfied when environmental covariates are present in the model.

We will use two genetic data sets presented by \cite{TOP1}, \cite{TOP2}, \cite{VO2max1} and \cite{VO2max2} to illustrate our method applied to real data. 

The paper is organized as follows. In Section \ref{sec:background} we present statistical background on the score test, and derive expressions for the score test covariance matrix, which is of importance for the subsequent work. Our proposed method is outlined and presented in detail in Section \ref{sec:multtest}, together with characteristics of our method. In Section \ref{sec:invest} real data and an artificial correlation structure are used to evaluate our proposed model and compare to other methods. Finally, we discuss and conclude in Sections~\ref{sec:discuss} and~6.

\section{Statistical background}\label{sec:background}

In this section, we present notation and details on the score test in generalized linear models. 

\subsection{Notation and data} \label{sec:notation}

We assume that data -- phenotype, $m$ genetic covariates and $d$ environmental covariates -- from $n$ independent individuals are available in a case--control, cohort or cross-sectional study. Let $\bm Y$ be an $n$-dimensional vector having the phenotype $Y_i$ of individual $i$ as its $i$th entry, $i=1$, \ldots, $n$. Let $\xe$ be an $n\times d$ matrix having environmental covariates (the first one being 1 to allow for an intercept in the model presented below) for individual $i$ as its $i$th row, and let $\xg$ be an $n\times m$ matrix having genetic covariates, or genotypes, for individual~$i$ as its $i$th row, each column corresponding to a genetic marker.

We assume that the genetic data are from common variant biallelic genetic markers with alleles $a$ and $A$, where $A$ is the minor allele. We will use the additive coding 0, 1, 2 for the genotypes $aa$, $aA$, and $AA$, respectively, in the genetic covariate matrix $\xg$, but other coding schemes are also possible. We denote the total design matrix $\x=(\xe\ \xg)$, which has the total covariate vector for individual $i$ as its $i$th row.

\subsection{Testing statistical hypotheses with the score test}\label{sec:score}

We assume that the relationship between the phenotype $\y$ and covariates $\x$ can be modelled by a generalized linear model (GLM) \citep{glm} with an $n$-dimensional vector $\bm\eta = \xe\be + \xg\bg = \x\bm\beta$ of linear predictors, where $\bm\beta=(\be^T\ \bg^T)^T$ is a $d+m$-dimensional parameter vector. Let $\eta_i$ be the $i$th entry of $\bm\eta$, and let $\bm\mu$ be the $n$-dimensional vector having $\mu_i=EY_i$ as its $i$th entry. We assume that the link function $g$ defined by $\eta_i=g(\mu_i)$ of the GLM is canonical, which implies that the log likelihood for individual~$i$ is $l_i=(Y_i\eta_i-b(\eta_i))/\phi_i+c(Y_i,\phi_i)$, where $b$ and $c$ are functions defining the exponential family of the phenotypes and $\phi_i$ the dispersion parameter. In our context $\phi_i=\phi$ will be equal for all observations. In general, $\mu_i=b'(\eta_i)$ and $\var\y[i]=\sigma^2_i=\phi b''(\eta_i)$. For $Y_i$ normally distributed, this reduces to $\sigma_i^2=\sigma^2=\phi$, and for $Y_i$ Bernoulli distributed,  $\sigma_i^2=\mu_i(1-\mu_i)$ with $\phi=1$.

The full $d+m$-dimensional score vector $\sum_{i=1}^n\nabla_{\!\bm\beta}l_i$ can then be calculated to be
\[
 \bm U=\frac1\phi\x^T(\y-\bm\mu),
\]
which is asymptotically normal with mean $\bm0$ and covariance matrix 
\[
 V=\frac1{\phi^2}\x^T\Lambda\x,
\]
where $\Lambda$ is the diagonal matrix having $\sigma_i^2$ as its $i$th entry.

Partition $\bm U$ into its environmental and genetic components, $\bm U^T=(\bm U_\text e^T\ \bm U_\text g^T)$. Since $\be$ are nuisance parameters and unknown, they are estimated by their maximum likelihood estimates under the null hypothesis of $\bg=\bm0$. In effect, $\bm\mu$ is to be replaced by $\hat{\bm\mu}_\text e$, the fitted values in a model with only environmental covariates $\xe$ present, giving the statistic
\begin{equation}
 \bm U_{\text g\mid\text e}=\frac1\phi\xg^T(\y-\hat{\bm\mu}_\text e).\label{u}
\end{equation}
Then $\bm U_{\text g\mid\text e}$ has the conditional distribution of $\bm U_\text g$ given $\bm U_\text e=\bm 0$, which is asymptotically normal with mean $\bm 0$ and covariance matrix
\begin{equation}
 V_{\text g\mid\text e}^{\vphantom1}=V_\text{gg}^{\vphantom1}-V_\text{ge}^{\vphantom1}V_\text{ee}^{-1}V_\text{eg}^{\vphantom1}
 =\frac{1}{\phi^2}\xg^T\big(\Lambda-\Lambda    \xe(\xe^T\Lambda\xe)^{-1}\xe^T \Lambda\big)\xg,\label{var}
\end{equation}
where $V_\text{ee}$, $V_\text{eg}$, $V_\text{ge}$ and $V_\text{gg}$ are the upper left $d\times d$, upper right $d\times m$, lower left $m\times d$ and lower right $m\times m$ submatrices of $V$, respectively \citep[see][]{Smyth2003}.

The score test statistic $\bm U_{\text g\mid\text e}^TV_{\text g\mid\text e}^{-1}\bm U_{\text g\mid\text e}$ with $\bg=\bm 0$ is asymptotically $\chi^2$ distributed with $m$ degrees of freedom when the complete null hypothesis $\bg=\bm0$ is true \citep[see][]{Smyth2003}. However, our interest lies not in the complete null hypothesis, but in the $m$ individual hypotheses $\bg[j]=0$ for each component $\bg[j]$ of $\bg$, \ $1\leq j\leq m$, against two-sided alternatives. We consider the standardized components of $\bm U_{\text g\mid\text e}$,
\begin{equation}
 T_j=\frac{\bm U_{\text g\mid\text e\,j}}{\sqrt{V_{\text g\mid\text e\,jj}}}, \label{eq:Tk}
\end{equation}
where $\bm U_{\text g\mid\text e\,j}$ denotes the $j$th entry of  $\bm U_{\text g\mid\text e}$ and $V_{\text g\mid\text e\,jk}$ the $jk$ entry of $V_{\text g\mid\text e}$. Under the null hypothesis $H_j\colon \bg[j]=0$, \ $T_j$ is asymptotically standard normally distributed, and $H_j$ will be rejected for large values of $\lvert T_j\rvert$. Under the complete null hypothesis, $\bg=\bm0$, the vector $\bm T=(T_1,T_2,\ldots,T_m)$ is asymptotically multivariate standard normally distributed with covariance matrix $R$, having
\begin{equation}
 \cov(T_j,T_k)=\frac{V_{\text g\mid\text e\,jk}}{\sqrt{V_{\text g\mid\text e\,jj}V_{\text g\mid\text e\,kk}}}, \label{eq:R}
\end{equation}
as its $jk$ entry, all evaluated at $\bg=\bm0$. Note that the dispersion parameter $\phi$ is cancelled from $\bm T$ and the covariances. However, the $\sigma_i^2$ of $\Lambda$ will have to be estimated.

\subsection{Special cases} \label{sec:corr}

We will now look at $\bm U_{\text g\mid\text e}$ and  $V_{\text g\mid\text e}$ for some special cases.

\subsubsection{No environmental covariates}\label{noenvcov}

If no evironmental covariates except the intercept are present in the GLM, then $\xe=\bm1$, the $n$-dimensional vector having all entries equal to 1, and $\Lambda=\sigma^2I$ under the null hypothesis, where $I$ is the $n\times n$ identity matrix. Then
\[
 U_{\text g\mid\text e}=\frac1\phi\xg^T\Big(I-\frac{1}{n}\bm1\bm1^T\Big)\bm Y\qquad\text{and}\qquad V_{\text g\mid\text e}=\frac{\sigma^2}{\phi^2}\xg^T\Big(I-\frac{1}{n}\bm1\bm1^T\Big)\xg,
\]
so that
\begin{equation}
T_j=\frac{\bm x_j^T(I-\frac{1}{n}\bm1\bm1^T)\bm Y}{\sigma\sqrt{\bm x_j^T(I-\frac{1}{n}\bm1\bm1^T)\bm x_j}},\quad\cov(T_j,T_k)=\frac{\bm x_j^T(I-\frac{1}{n}\bm1\bm1^T)\bm x_k}{\sqrt{\bm x_j^T(I-\frac{1}{n}\bm1\bm1^T)\bm x_j}\sqrt{\bm x_k^T(I-\frac{1}{n}\bm1\bm1^T)\bm x_k}},\label{Tk-noenv}
\end{equation}
where $\bm x_j$ is the $j$th column of $\xg$, \ $1\leq j\leq m$, \ $1\leq k\leq m$. So $T_j$, the score test statistic for testing $\bg[j]=0$, is $\sqrt n$ times the Pearson correlation between $\bm x_j$ and $\bm Y$ when $\sigma^2=\var Y_i$ is replaced by the estimate $\bm Y^T(I-\frac{1}{n}\bm1\bm1^T)\bm Y/n$, and $\cov(T_j,T_k)$ is the sample correlation between $\bm x_j$ and $\bm x_k$. Thus, for a GLM without adjustment for environmental covariates, the correlation between the score test statistics can be estimated by estimating the genotype correlation. The genotype correlation estimates twice the composite linkage disequilibrium if the genotypes are coded 0, 1,~2 \citep{Weir2008}.

\subsubsection{Uncorrelated environmental and genetic covariates}\label{uncorrenvgen}

Two $n$-dimensional vectors $\bm X_1$ and $\bm X_2$ of observations have zero Pearson correlation if their centered observations are orthogonal,
\[
 0=(\bm X_1-\bar X_1\bm1)^T(\bm X_2-\bar X_2\bm1)=\bm X_1^T\Big(I-\frac1n\bm1\bm1^T\Big)\bm X_2.
\]
If $X_1$ and $X_2$ are two matrices, then near zero Pearson correlation of each combination of a column of $X_1$ and a column of $X_2$ can be written compactly as
\begin{equation}
 X_1^T\Big(I-\frac1n\bm1\bm1^T\Big)X_2\approx\bm0,\qquad\text{or}\qquad
   X_1^TX_2\approx\frac1nX_1^T\bm1\bm1^TX_2.\label{centeredorth}
\end{equation}

If we consider genetic and environmental covariates to be random variables, and all pairs of an environmental and a genetic covariate to be independent, we would expect~\eqref{centeredorth} to  hold for all $X_1$ having columns that are functions of genetic covariates and $X_2$ having columns that are functions of environmental covariates. In particular, we consider $X_1=\xg$ and $X_2=\Lambda\xe$. Since $\Lambda$ is a function of environmental covariates only under the null hypothesis, so is $X_2$. By~\eqref{centeredorth}, $\xg^T\Lambda\xe\approx\frac1n\xg^T\bm1\bm1^T\Lambda\xe$, Then, from~\eqref{var},
\begin{align*}
 \phi^2V_{\text g\mid\text e}
  &\approx\xg^T\Lambda\xg
   -\frac1{n^2}\xg^T\bm1\bm1^T\Lambda\xe(\xe^T\Lambda\xe)^{-1}\xe^T \Lambda\bm1\bm1^T\xg\\
  &=\xg^T\Lambda\xg
   -\frac1{n^2}\xg^T\bm1\bm1^T\Lambda^{1/2}H\Lambda^{1/2}\bm1\bm1^T\xg,
\end{align*}
where $H=\Lambda^{1/2}\xe(\xe^T\Lambda\xe)^{-1}\xe^T\Lambda^{1/2}$ will project onto the column space of $\Lambda^{1/2}\xe$. Since $\bm1$ is a column (the intercept) of $\xe$, \ $\Lambda^{1/2}\bm1$ is in the column space of $\Lambda^{1/2}\xe$, so that $H\Lambda^{1/2}\bm1=\Lambda^{1/2}\bm1$, and
\[
 \phi^2V_{\text g\mid\text e}
  \approx\xg^T\Lambda\xg
   -\frac1{n^2}(\tr\Lambda)\xg^T\bm1\bm1^T\xg.
\]

We now turn to the term $\xg^T\Lambda\xg$. Its $(j,k)$ entry is $\bm X_1^T\Lambda\bm1$, where $\bm X_1$ is the vector consisting of the entry-wise products of the $j$th and the $k$th column of $\xg$. Letting $\bm X_2=\Lambda\bm1$, by~\eqref{centeredorth}, independence of environmental and genetic covariates yields $\bm X_1^T\Lambda\bm1\approx\frac1n\bm X_1^T\bm1\bm1^T\Lambda\bm1=\frac1n(\tr\Lambda)\bm X_1^T\bm1$, which is the $(j,k)$ entry of $\frac1n(\tr\Lambda)\xg^T\xg$. Thus $\xg^T\Lambda\xg\approx\frac1n(\tr\Lambda)\xg^T\xg$, and we have
\[
 V_{\text g\mid\text e} \approx\frac{\tr\Lambda}{n\phi^2}\xg^T\Big(I-\frac1n\bm1\bm1^T\Big)\xg,
\]
which is the same expression as in the case of no environmental covariates with the exception that the common variance $\sigma^2$ of the responses is replaced by their average variance $\tr\Lambda/n=\frac1n\sum_{i=1}^n\sigma_i^2$, where the $\sigma_i^2$ are defined by the environmental covariates. The conclusion is that, if environmental and genetic covariates are uncorrelated, correlations of the score vector under the null hypothesis can be estimated more easily by estimating only correlations between genetic covariates instead 

\subsubsection{The normal model}

For $Y_i$ normally distributed, $\Lambda=\sigma^2I$, where $I$ is the $n\times n$ identity matrix. The score vector can then be written
\[
  \bm U_{\text g \mid\text e}=\frac1{\sigma^2} \xg^T (I-H)\y,
\]
and \eqref{var} reduces to
\[
  V_{\text g\mid\text e}=\frac{1}{\sigma^2}\xg^T(I-H)\xg,
\]
where $H=\xe(\xe^T\xe)^{-1}\xe^T$ is the idempotent matrix projecting onto the column space of $\xe$. Then $I-H$ is the idempotent matrix projecting onto the orthogonal complement of the column space of $\xe$, and $(I-H)\bm Y$ are the residuals when fitting the multiple linear model with only the environmental covariates present. Note that $\sigma^2$ enters into the test statistics $T_j$~\eqref{eq:Tk}, and needs to be replaced by an estimate; we have used the residual sum of squares of a fitted model with only environmental covariates present (the null hypothesis), divided by $n-d$.

\subsubsection{The logistic model}

For $Y_i$ Bernoulli distributed, $\phi=1$ and the $\sigma_i^2$ of $\Lambda$ are estimated by $\hat \mu_{\text ei}(1-\hat \mu_{\text ei})$, where $\hat \mu_{\text ei}$ are the fitted values under the null hypothesis with only environmental covariates. Inference about $\bg$ is valid also if data are collected in a case--control study since the canonical (logit) link is used \citep[pp. 170--171]{agresti2002categorical}.

In the special case of no environmental covariates, that is, $\xe=\bm1$, each score test statistic, $T_j$ \eqref{Tk-noenv}, is equal to the Cochran--Armitage trend test \citep{Armitage1955,Cochran1954} statistic,
\[
 \frac{\sum_{i=0}^2s_i(n_2x_i-n_1y_i)}
  {\sqrt{n_1n_2\big(\sum_{i=0}^2s_i^2m_i-\frac1n(\sum_{i=0}^2s_im_i)^2\big)}},
\]
where $s_i$ are the possible values of the genetic covariates, $n_1$ and $n_2$ the number of 0 and 1 phenotypes $Y_i$, respectively, $x_i$ the number of observations having phenotype 1 and genotype $s_i$ at marker $k$, \ $y_i$ the number of observations having phenotype 0 and genotype $s_i$, and $m_i=x_i+y_i$. The Cochran--Armitage test is used in disease--genotype association testing with scores $(s_0,s_1,s_2)=(0,s,1)$ \citep{sasieni1997genotypes,slager2001case}, for example with $s=\frac12$ for an additive genetic model.

\section{Familywise error rate control and approximations}\label{sec:multtest}

We now turn to the topic of how to control the familywise error rate (FWER) by intersection approximations, and then apply this to our situation.

\subsection{Multiple hypothesis familywise error rate control}\label{sec:fwer}

We have a collection of $m$ null hypotheses, $H_k\colon \bg[k]=0$ (no association between phenotype and genotype at marker $k$), $1\leq k\leq m$, against two-sided alternatives. We will present a method for multiple testing correction that controls the FWER -- the probability of making at least one type~I error. We adopt the notation of \cite{MoskvinaSchmidt2008}, and denote by $O_k$ the event that the null hypothesis $H_k$ is not rejected, and by $\bar O_k$ its complement, $1\leq k\leq m$. Then, if all $m$ null hypotheses are true,
\begin{equation}
 \text{FWER} = P(\bar O_1 \cup \cdots \cup \bar O_m) = 1-P(O_1 \cap \cdots \cap O_m). \label{eq:FWER}
\end{equation}
In our case, $O_k$ is an event of the form $|T_k| < c$, where $T_k$ is the test statistic of \eqref{eq:Tk}. We will consider single-step multiple testing methods, and choose the same cut-off $c$ for each $k$. We denote by $\al=2\Phi(-c)=P(\bar O_k)$, the asymptotic probability of false rejection of $H_k$, where $\Phi$ is the univariate standard normal cumulative distribution function. When the joint distribution of the test statistics is known under the complete null hypothesis, or can be estimated, FWER control at the $\alpha$ significance level can be achieved by solving the inequality $\text{FWER} \leq \alpha$ for $\al$, based on either the union or intersection formulation of~\eqref{eq:FWER}. When $m$ is large, this involves evaluating high dimensional integrals over the acceptance or rejection regions, which is suggested by \cite{ConneelyBoehnke2007}.

To avoid evalulating these costly integrals, we may instead control FWER by considering bounds based on~\eqref{eq:FWER}. For example, the Bonferroni method is based on the Boole inequality applied to the union formulation of~\eqref{eq:FWER},
\[
 \text{FWER} = P(\bar O_1 \cup \cdots \cup \bar O_m) \leq \sum_{k=1}^m P(\bar O_k)=\sum_{k=1}^m \al=m\al,
\]
from which it is seen that a local significance level of $\al=\alpha/m$ guarantees $\text{FWER}\leq\alpha$.

When the FWER is calculated under the complete null hypothesis, so-called weak FWER control is achieved. However, in our situation, subset pivotality is satisfied, meaning that the distribution of any subvector $(T_k)_{k\in K}$ is identical under $\bigcap_{k\in K}H_k$ and under the complete null hypothesis $\bigcap_{k=1}^mH_k$, for all subsets $K\subseteq\{1,2,\ldots, m\}$. In particular, a subvector of $\bm U_{\text g\mid\text{e}}$~\eqref{u} and a submatrix of $V_{\text g\mid\text{e}}$~\eqref{var} corresponding to $K$ only involves genetic covariates corresponding to~$K$. Then strong FWER control is achieved, meaning that $\text{FWER}\leq\alpha$ regardless of which null hypotheses are true \citep{WestfallYoung1993,westfall2008multiple}. 

The focus in this work will be on the intersection formulation of~\eqref{eq:FWER}. Background theory will be given next and new application in \ref{sec:FWERkapprox}.

\subsection{Intersection approximations}\label{sec:fwerk}

Following \cite{glazjohnson}, we define $k$th order product-type approximations to $P(O_1\cap\nobreak\cdots\cap O_m)$ by
\begin{equation}
 \gamma_k=P(O_1\cap\cdots\cap O_k)\prod_{j=k+1}^mP(O_j\mid O_{j-k+1}\cap\cdots\cap O_{j-1})
  =\frac{\prod_{j=k}^{m}P(O_{j-k+1}\cap\cdots\cap O_j)}{\prod_{j=k+1}^{m}P(O_{j-k+1}\cap\cdots\cap O_{j-1})},
\label{gammak}
\end{equation}
$1\leq k\leq m$, where probabilities are evaluated under the complete null hypothesis. This is similar to the usual multiplicative rule for the probability of intersection of events applied to $\gamma_m=P(O_1\cap\nobreak\cdots\cap O_m)$, but with dimension of distributions limited to $k$. The idea is that the $\gamma_k$ should constitute increasingly better approximations of $\gamma_m$ as $k$ increases, and that calculation of $\gamma_k$ is less costly than calculation of $\gamma_m$ when $k<m$, since only $k$-variate distributions are involved in $\gamma_k$.

\sloppy
Note that the approximations depend on the order of the components of $\bm T=(T_1,\ldots T_m)$. We have used the order in which the $m$ markers are positioned along the genome, assuming that the largest correlations occur between close markers.

\fussy
In our case, $\gamma_1=\prod_{j=1}^m P(\lvert T_j\rvert<c)=(1-\al)^m$ and $\gamma_m=P(\lvert T_1\rvert<c,\ldots,\lvert T_m\rvert<\nobreak c)=1-\text{FWER}$. Since $\bm T$ is asymptotically multivariate normally distributed with mean~$\bm0$ under the complete null hypothesis, $\gamma_1\leq\gamma_m$ asymptotically \citep{Sidak1967}. Choosing $\al$ such that $\text{FWER}=1-\gamma_m\leq1-\gamma_1=1-(1-\al)^m=\alpha$ keeps FWER at the $\alpha$ level. It is well known that the $\al$ found by this method, the \Sidak method, is slightly larger than the $\al$ found by the Bonferroni method, thus the \Sidak method will give slightly higher power.

We have seen that in our case, $\gamma_1\leq\gamma_m$, meaning that the \Sidak method can safely be used. If $\gamma_k\leq\gamma_m$, then $\text{FWER}=1-\gamma_m\leq1-\gamma_k=\alpha$ can be used to control FWER by solving the last equation for $\al$ (choosing the greatest solution if not unique -- we have, however, never observed a $\gamma_k$ that is not monotonically decreasing in $\al$). If $\gamma_k\leq\gamma_l$, then continuity of $\gamma_k$ and of $\gamma_l$ as functions of $\al$ implies that the $\al$ making $1-\gamma_l=\alpha$ is no less than the $\al$ making $1-\gamma_k=\alpha$, so that the power obtained by the $l$th approximation is no less than the power obtained by the $k$th approximation.

\sloppy
The ideal property $\gamma_1\leq\gamma_2\leq\cdots\leq\gamma_k\leq\gamma_m$ for all $\al$ is ensured if $\lvert\bm T\rvert=(\lvert T_1\rvert,\ldots,\lvert T_1\rvert)$ is monotonically sub-Markovian of order $k$ ($\text{MSM}_k$) with respect to $(-\infty,c)^k$ for all $c$, \ $2 \leq k \leq m-1$, as defined by \cite{Block1992}. Unfortunately, our $\lvert\bm T\rvert$ is not $\text{MSM}_{m-1}$. It is possible to construct a trivariate normal distribution with mean $\bm0$ such that $\gamma_1<\gamma_3<\gamma_2$ for some~$\al$. However, the violations of $\text{MSM}_{m-1}$ we have observed have been very small and only for restricted ranges of $\al$, and only for carefully constructed covariance matrices. We have not observed violations for covariance matrices estimated from real data, and will therefore proceed to apply $\gamma_2$ and $\gamma_3$ as better approximations to $\gamma_m$ than $\gamma_1$ (the latter giving \Sidak cutoffs). A summary of concepts of positive dependence, like $\text{MSM}$, was given by \citet[pp.~58--61]{Dickhaus2014}.

\fussy
\subsection{Controlling FWER using $k$th order approximation for score tests}\label{sec:FWERkapprox}

As we have seen, the vector $\bm T$ of score test statistics is under the complete null hypothesis asymptotically standard multivariate normal with covariance matrix $R$ \eqref{eq:R}. We denote by $O_j$ the event $\lvert T_j\rvert<c$ of non-rejection of $H_j$, which has probability $P(O_j)=1-\al$ under the null hypothesis, with $\al=2\Phi(-c)$. We will detail how to find $\al$ given by the second order approximation, $\gamma_2$: Denote by $r_j$ the $(j-1,j)$ entry of $R$. Then
\[
 P(O_{j-1}\cap O_j)= 1-\al-\sqrt{\frac{2}{\pi}} \int_{-c}^{c}\!\!e^{-x^2/2}\,\Phi\Biggl(\frac{r_j x-c}{\sqrt{1-r_{\vphantom i\smash j}^2}}\Biggr)dx,
\]
giving
\begin{align}
 \gamma_2&=P(O_1\cap O_2)\prod_{j=3}^mP(O_j\mid O_{j-1})
  =\frac{\prod_{j=2}^m P(O_{j-1} \cap O_j)}{\prod_{j=3}^{m} P(O_{j-1})}\nonumber\\
 &= \frac{\prod_{j=2}^{m}\Bigl(1-\al-\sqrt{\frac{2}{\pi}}\int_{-c}^{c}e^{-x^2/2}\,\Phi\Bigl(\frac{r_{j}x-c}{\sqrt{1-r_{\vphantom i\smash j}^2}}\Bigr)dx\Bigr)}{(1-\al)^{m-2}}\nonumber\\
 &= (1-\al)\prod_{j=2}^{m}\Biggl(1-\sqrt{\frac{2}{\pi}}\frac{1}{1-\al}\int_{-c}^{c}\!\!e^{-x^2/2}\,\Phi\Biggr(\frac{r_{j}x-c}{\sqrt{1-r_{\vphantom i\smash j}^2}}\Biggr)dx\Biggr).
\label{eq:gamma2}
\end{align}
For a desired upper bound $\alpha$ on FWER, the equation $1-\gamma_2=\alpha$ is solved with respect to $\al$, which can be done numerically using for example a bisection algorithm. Note that $\al$ enters into $c=-\Phi^{-1}(\al/2)$.

We can control FWER by higher-order approximations by solving the equation $1-\gamma_k=\alpha$ for $\al$ in a similar way, which we will henceforth refer to as order $k$ FWER approximation. By~\eqref{gammak}, $\gamma_k$ can be written as a ratio of products of $k$-dimensional and products of $k-1$-dimensional multivariate normal integrals. Good numerical methods for calculating multivariate normal integrals exist for small dimensions \citep{genz2009computation}. We will illustrate using $k=2$ and $k=3$ for real data in Section \ref{sec:invest}.

The procedure to find $\al$ does not depend on the exact form of the test statistic, only that the vector $(T_1,\ldots,T_m)$ of test statistics is asymptotically standard multivariate normal under the complete null hypothesis and $\lvert T_j\rvert\geq c$ leads to rejection. In particular, \eqref{eq:gamma2} is identical to what was found by \cite{MoskvinaSchmidt2008} for an allelic test and correlations given by linkage disequilibria.

In practice, instead of calculating $\al$, it may be preferable to calculate FWER-adjusted $p$-values: Replace $\al$ with $p$, the unadjusted $p$-value for an individual test, in the calculation of $\gamma_k$. Then $1-\gamma_k$ is an FWER-adjusted $p$-value for the test, in the sense that if $1-\gamma_k\leq\alpha$ (rejection based on adjusted $p$-value), then $p\leq\al$ (rejection based on local significance level).

\subsection{FWER control with independent blocks}\label{sec:blocks}

Genetic markers are distributed along the chromosomes and a common assumption is independence of genetic markers from different chromosomes.

As we have seen in Section \ref{sec:corr}, if the genetic markers are independent and no environmental covariates that are correlated with the genetic markers are included, the score test statistics for these markers would also be independent. Within a chromosome, genetic markers can belong to different haplotype blocks, being highly correlated within a block and independent or nearly independent between the blocks \citep{biologi}.

Assume that the $m$ markers to be tested, and thus $\{O_1,\ldots,O_m\}$, can be partitioned into $b$ independent blocks, $\{O_1,\ldots,O_{m_1}\}$, \ $\{O_{m_1+1},\ldots,O_{m_2}\}$, \ldots, $\{O_{m_{b-1}+1},\ldots,O_m\}$,  so that $O_{j_1}$ and $O_{j_2}$ are independent if they belong to different blocks. Let $\gamma_k^{(l)}$ be the $k$th order approximation given by~\eqref{gammak} for the intersection of the events belonging to the $l$th block, \ $1\leq l\leq b$, and let $\gamma_k$ be the overall $k$th order approximation. Then it is easy to verify that $\gamma_k=\prod_{l=1}^b\gamma_k^{(l)}$.

\subsection{The effective number of independent tests}\label{sec:Meff}

The concept of an effective number of independent tests, $\Meff$, in multiple
testing problems has been described and discussed by many authors, including \cite{nyholt}, \cite{gao1}, \cite{MoskvinaSchmidt2008}, \cite{LiJi2005}, \cite{Galwey2009} and \cite{ChenLiu2011}. All except \cite{MoskvinaSchmidt2008} first estimate $\Meff$, and then use $\Meff$ in place of $m$ in the \Sidak formula to calculate $\al=1-(1-\alpha)^{1/\Meff}$. An alternative formulation using the Bonferroni formula also exists.

None of these methods use the concept of FWER in the derivation of $\Meff$, and there is no mathematical justification that FWER is controlled. All methods start with the linkage disequilibrium or composite linkage disequilibrium matrix, and there is no mention of the dependence of the $\Meff$ estimate on the test statistics used for the hypothesis tests.

The method of \cite{MoskvinaSchmidt2008} is based on an allelic test and controls the FWER using second order intersection approximations. As for our method, the main output of their method is an estimate of $\al$. The above \Sidak formula can then be used to define $\Meff=\ln(1-\alpha)/\ln(1-\al)$. Note that $\Meff$ depends on both $\al$ and the FWER threshold~$\alpha$. We will not consider $\Meff$ further in this article. 

\subsection{The maxT permutation method}\label{maxT}

We will compare the local significance level $\al$, as calculated by the FWER approximation method presented in section~\ref{sec:FWERkapprox}, with the \cite{WestfallYoung1993} maxT permutation method, and give a brief review of the latter.

The FWER is the probability that at least one of the $m$ null hypotheses is falsely rejected, which can be formulated as $P(\max_j\lvert T_j\rvert\geq c)=1-\gamma_m$ under the complete null hypothesis. In the maxT method, the critical value $c$ is found empirically by permutation of the response variable in order to generate a sample from the distribution of the $\max_j\lvert T_j\rvert$ statistic. If the FWER is to be controlled at the $\alpha$ level and $b$ permutations are made, $c$ is estimated by the $(1-\alpha)b$th order statistic of the $\max_j\lvert T_j\rvert$ (the $(1-\alpha)b$th smallest value), which is an estimate of the $1-\alpha$ quantile of $\max_j\lvert T_j\rvert$. The probability that the $k$th order statistic of a random sample of size $b$ is greater than the $1-\alpha$ quantile is equal to the binomial cumulative distribution function with parameters $b$ and $1-\alpha$ evaluated at $k-1$, which can be used to construct a confidence interval for $c$ (\citeauthor{thompson1936confidence}, \citeyear{thompson1936confidence}, see e.g. \citeauthor{conover1980practical}, \citeyear{conover1980practical}, p. 114). A confidence interval for $\al$ is obtained by transforming the bounds via $\al=2\Phi(-c)$.

The success of the permutation method relies on the exchangeablity of the data, which in general does not hold for regression problems \citep{Commenges2003}.  In our GLM the responses $\bm Y$ are in general not exchangeable since their expected values are not equal when environmental covariates (which may not be independent of the genetic covariates) are present. Without environmental covariates (only intercept) the responses are exchangeable and permutation of $\bm{Y}$ gives FWER control. With discrete environmental covariates permutation can be done in a stratified manner \citep{Solari2014}. When the exchangeability assumption is not satisfied, there is no standard solution to how permutation testing can be performed. Asymptotic or second moment exchangeability may be obtained by different transformations of the data, but comparison with these methods is beyond the scope of this paper. 

\section{Power and efficiency of the FWER approximation method}\label{sec:invest}

We will compare the local significance level $\al$, as calculated by the FWER approximation method presented in the previous section, with $\al$ of the Bonferroni method and of the \cite{WestfallYoung1993} maxT permutation method. We proceed to compare the $\al$ calculated by FWER approximation in two cases were the ``true'' $\al$ based on the entire joint distribution can be calculated; one artificial and one based on data.

\subsection{Illustration of methods: TOP and $\text{VO}_2$-max data}\label{sec:topvo2}

Our two data sets (referred to as TOP and $\text{VO}_2$-max) are of limited sample size, and our aim is to use the data to investigate the correlation structure of GWA-data and the effect this has on the estimation of the local significance level. We assume that our findings will hold in data sets with larger sample sizes. An increase in sample size will give more precise estimates of the score test statistics correlations, but the estimation of the local significance level is mainly dependent on score test statistics correlations under study (not the sample size).

The TOP data set is a case--control GWA data set, in which case is schizophrenia or bipolar disorder. The data set was collected with the aim to detect single-nucleotide polymorphisms (SNPs) associated with the schizophrenia or bipolar disorder \citep{TOP1,TOP2}. The preprocessed TOP GWA data contain genetic information on 672972 SNPs (Affymetrix Genome-Wide Human SNP Array 6.0) for 1148 cases and 420 controls. Our dataset included individuals sampled until March 2013, and therefore the sample size is larger than in the cited papers. Preprosessing of the data was done as described in \cite{TOP1} and \cite{TOP2}.

Genotype--phenotype association was assessed by fitting a logistic regression without any environmental covariates, so that score test correlations equal genotype correlations (Section~\ref{noenvcov}).
 
The $\text{VO}_2$-max data set comes from a cross-sectional GWA study \citep{VO2max1, VO2max2}, in which the aim was to find SNPs associated with maximum oxygen uptake. The preprocessed $\text{VO}_2$-max GWA data consist of 123497 SNPs \citep[Illumina Cardio-MetaboChip,][]{VO2chip} for 2802 individuals. The $\text{VO}_2$-max data were analysed using a normal linear regression model, including age, sex and physical activity score as covariates. For both datasets, some genotype data were missing. In the TOP data, mean imputation was done for $0.04\%$ of the genotypes, and in the $\text{VO}_2$-max data for $0.7\%$ of the genotypes. 

For the $\text{VO}_2$ data, the local significance level controlling the FWER at level 0.05 was lowest for the Bonferroni method, slightly higher for the order 1 approximation (the \v Sid\'ak method), and further increasing through the order 2 and~3 FWER approximations (Table~\ref{tab:resultater}).

\begin{table}
\begin{center}
\begin{tabular}{lccccc}\hline
 & \multicolumn2c{TOP} && \multicolumn2c{$\text{VO}_2$-max}\\ \cline{2-3}\cline{5-6}
Method & $10^8\al\vphantom{\big)}$ & Ratio & & $10^7\al$ & Ratio \\
\hline
Bonferroni  & 7.43 & 1.00 && 4.05 & 1.00\\
Order 1 (\v Sid\'ak)   & 7.62 & 1.03 && 4.15 & 1.02\\
Order 2 & 8.62 & 1.16 && 4.70 & 1.16 \\
Order 3 & 9.07 & 1.22 && 5.02 & 1.24 \\
\hline
\end{tabular}
\caption{Local significance level $\al$ calculated by the Bonferroni method and by order 1--3 FWER approximations for the TOP and $\text{VO}_2$-max data, controlling the FWER at level $0.05$, 
and ratio of $\al$ to Bonferroni $\al$.}
\label{tab:resultater}
\end{center}
\end{table}

For the TOP data, since no environmental covariates are included, permutation of the binary response vector is feasible
(the exchangeability assumption is satisfied), and the maxT method can be used to estimate the local significance level controlling FWER at level $0.05$. Permutation of the responses, followed by calculation of the maximal score test statistics over the whole genome, is a time consuming task, and we will only present results on two of the smallest chromosomes (chromosome 21 and 22):

The local significance level controlling the FWER at level 0.05 was, as for the $\text{VO}_2$ data, lowest for the Bonferroni method and increasing through order 1--3 FWER approximations (Table \ref{tab:TOPres}). The highest level was obtained for the maxT method, and also the lower bound of the 95\% confidence interval for $\al$ of maxT was greater than the order 3 FWER approximation.

On a $4\times6$-core Xeon 2.67 GHz computer (Intel CPU) running Linux (Ubuntu 14.0) using one thread, the analyses on chromosome 22 took 85 hours for maxT, 20 minutes for order 3 FWER and 10 seconds for order 2 FWER approximation.

\begin{figure}
\includegraphics[width=0.49\textwidth]{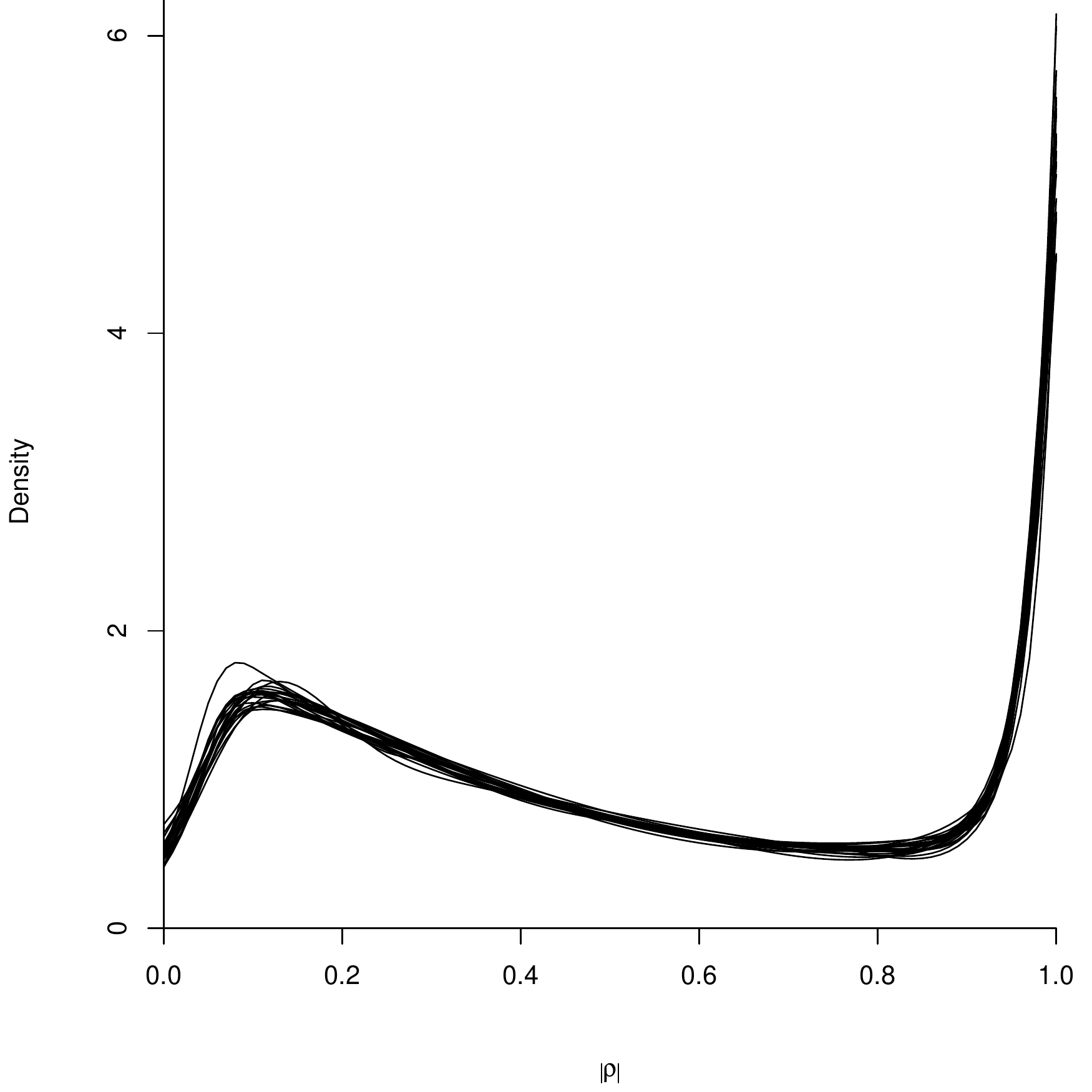}\hfill
\includegraphics[width=0.49\textwidth]{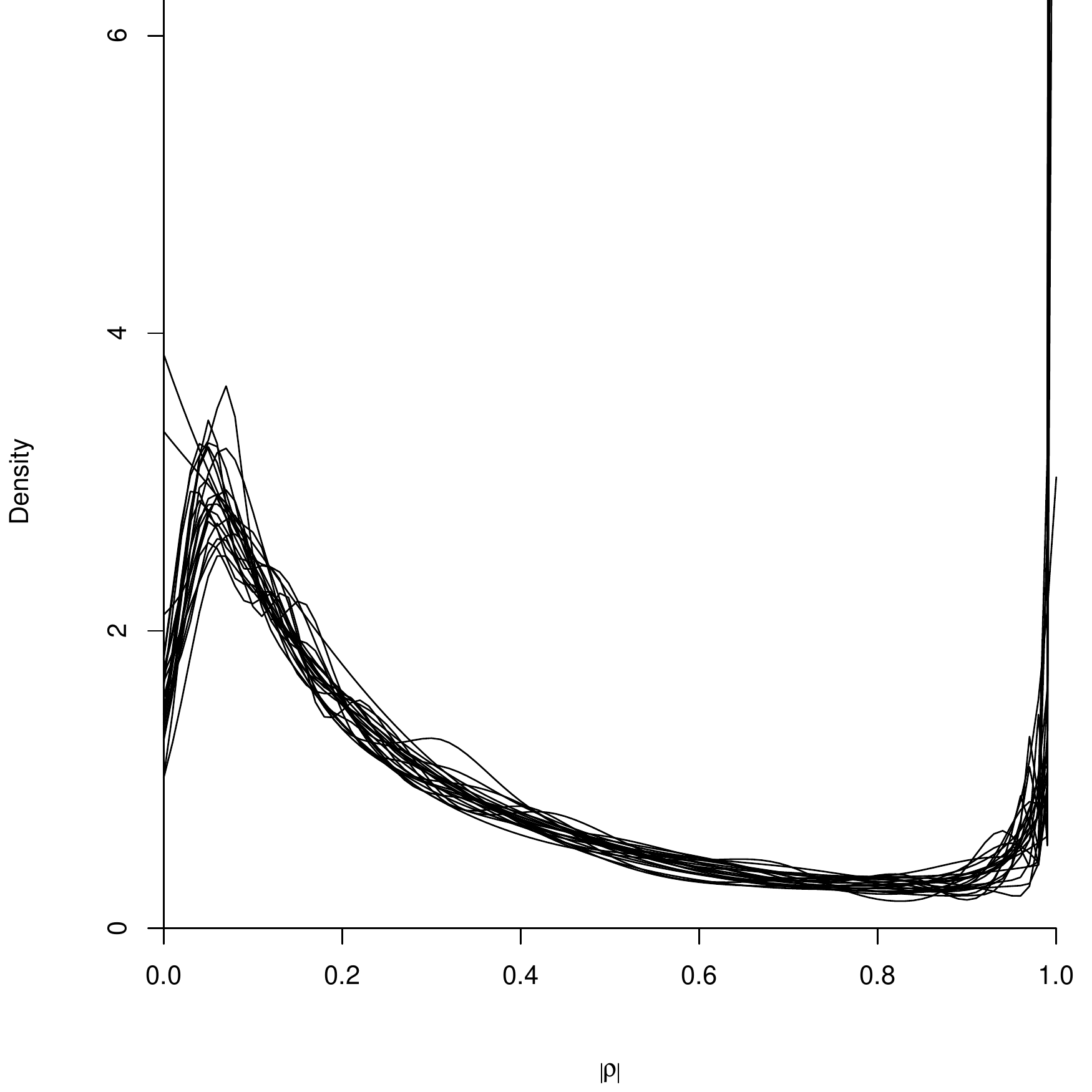}
\caption{Smoothed frequency distributions of absolute value of the estimated genotype correlations between neighbouring SNPs on each chromosome, one line per chromosome. TOP data (left) and $\text{VO}_2$-max data (right). The plots are logspline density estimates \citep{stone1997polynomial}, implemented by the R logspline package \citep{logsplineR}.}
\label{fig:densTOPandVO2}
\end{figure}

Smoothed frequency distributions of the estimated correlations between neighbouring SNPs along chromosomes are very similar across chromosomes (Figure \ref{fig:densTOPandVO2}), and therefore, we would expect that the trends for chromosome 21 and 22 can be extended to the other chromosomes and to the whole genome.

\begin{table}
\begin{center}
\begin{tabular}{lccccc}\hline
 & \multicolumn5c{Chromosome}\\ \cline{2-6}
 & \multicolumn2c{21} && \multicolumn2c{22}\\ \cline{2-3}\cline{5-6}
Method & $10^6\al\vphantom{\big)}$ & Ratio && $10^6\al$ & Ratio\\
\hline
Bonferroni & 5.10 & 1.00 && 5.57 & 1.00\\
Order 1 (\v Sid\'ak) & 5.23 & 1.03 && 5.72 & 1.03\\
Order 2 & 6.09 &1.19 && 6.46 & 1.16\\
Order 3 & 6.57 & 1.29 && 6.93 &1.24\\
maxT, lower & 7.34 & 1.44 && 7.68 & 1.38\\
maxT & 7.44 &1.46 && 7.79 &1.40\\
maxT, upper & 7.55 & 1.48 && 7.91 & 1.42\\ \hline
\end{tabular}
\end{center}
\caption{Local significance level $\al$ for the TOP data calculated by the Bonferroni method and by order 1--3 FWER approximations, and estimated by the maxT method, controlling FWER at level $0.05$, and ratio of $\al$ to Bonferroni $\al$. Chromosome 21 contained $9802$ SNPs and chromosome~22 contained $8970$ SNPs. The number of permutations for the maxT method was 500000. The lower and upper values for maxT are bounds of a 95\% confidence interval for $\al$ (see Section~\ref{maxT}).}
\label{tab:TOPres}
\end{table}

\subsection{Correlation structure and local significance level}\label{sec:alphaksim}

Consider 100 markers and a multivariate normal test statistic $\bm T$ having an AR1 correlation structure, that is, all entries on the main diagonal of the $100\times100$ correlation matrix are equal to~1, on the sub- and superdiagonal $\rho$, on the next diagonals $\rho^2$, and so on. We investigated the effect of positive $\rho$ on the local significance level $\al$ found by order 1--4 FWER approximations to control FWER at the 0.05 level. Also, the ``true'' $\al$ was calculated without approximation (that is, based on $\gamma_{100}$; see Section~\ref{sec:fwerk}), using the \texttt{pmvnorm} function of the R \citep{R} package mvtnorm \citep{mvtnormR} using the Genz--Bretz algorithm \citep{Genz1992,Genz1993,GenzBretz2002}. The \texttt{pmvnorm} function can calculate multivariate normal probabilities with some accuracy for dimensions up to 1000.

The inverse of an AR1 correlation matrix contains only negative off-diagonal entries, which ensures a property called  $\text{MTP}_2$ \citep{karlin1981total} for the density of $\lvert\bm T\rvert$, which implies that the product-type approximations $\gamma_k$ of Section~\ref{sec:fwerk} are non-decreasing in $k$ \citep{glazjohnson}, making the $\al$ of the order $k$ FWER approximations non-decreasing in $k$.

\begin{figure}
\centering\includegraphics[width=0.49\textwidth]{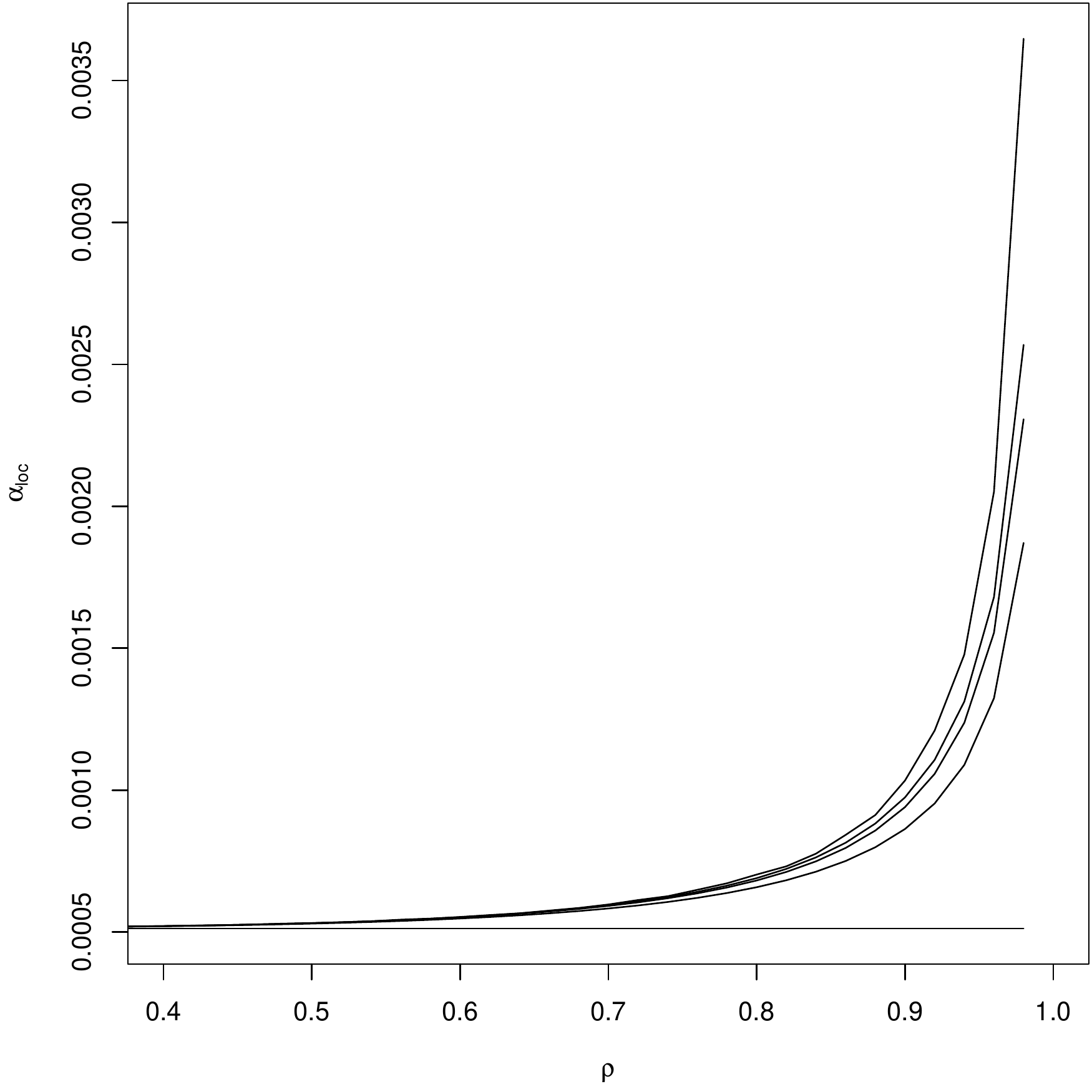}
\caption{Local significance level $\al$ for order 1--4 FWER approximations and $\al$ based on true joint distribution of test statistic as a function of the parameter $\rho$ of an AR1 correlation matrix for 100 markers. The horizontal line corresponds to \v Sid\'ak correction (order 1 FWER approximation), then $\al$ is increasing with the order of the approximation (order 2--4; the three curves in the middle). The uppermost curve shows $\al$ based on the true joint distribution.}
\label{fig:ar1}
\end{figure}

The effect of $\rho$ on $\al$ was small for $\rho<0.4$ (Figure \ref{fig:ar1}), so for the 100 markers considered, there would be no gain in using FWER approximation or even the true joint distribution of $\bm T$ instead of \v Sid\'ak this case. For larger $\rho$, order 2 FWER approximation provides an improvement compared to \v Sid\'ak. The increase in $\al$ from \v Sid\'ak to order 2 FWER approximation was greater than the difference between higher orders.

To assess the order $k$ FWER approximation method for a more realistic correlation structure, we considered the empirical correlation matrix for the first 1000 markers on chromosome 22 of the TOP data. The order 1--4 approximations to control FWER at the 0.05 level gave an $\al$ of 5.1 (\v Sid\'ak), 5.8, 6.2 and $6.4\cdot10^{-5}$, respectively, whereas the $\al$ calculated without approximation using the Genz--Bretz algorithm was $7.3\cdot10^{-5}$.

\section{Discussion}\label{sec:discuss}

We have presented the order $k$ FWER approximation method for estimating the local significance level $\al$ used to control FWER in a GWA study.  Our method takes the estimated correlation structure between the test statistics into account, and is applicable when environmental covariates are present. The relation between the phenotype response and the genetic and environmental covariates can be modelled by any generalized linear model (using the canonical link); in particular, both models with discrete and models with continuous phenotypes are allowed. We have applied the method to common genetic variants, but it can also be used for rare variants. However, since rare variants are less correlated than common variants, we expect the increase in $\al$ from the \Sidak method to be less than when analyzing common variants. 

The order $k$ FWER approximation is based on conditioning on the previous $k-1$ neighbouring markers along the chromosome. A sufficient condition to have non-decreasing local significance levels when the order of the FWER approximation increases from 1 to $k$, and that the order $k$ order approximation gives valid FWER control, is that the test statistic has the $\text{MSM}_k$ property. Even $\text{MSM}_2$ and $\text{MSM}_3$ are difficult to verify for our test statistic with GWA data, but it is reasonable to assume that they are satisfied (Section \ref{sec:fwerk}).

Population substructure can be associated with both the genotype and phenotype and is therefore a possible confounding factor in GWA studies. Population substructure can be adjusted for in the analysis using principal components of the covariance matrix of the individuals \citep{Price2006} as covariates. In both the TOP data and the VO$_2$-max data, related individuals were removed in the preprocessing of the data, and no adjustment for population structure was done in our analysis. 

The AR1 correlation structure (Section~\ref{sec:alphaksim}) might not be a realistic model for genotype correlations, but the calculations nevertheless show potential for a significant improvement over the \v Sid\'ak method by applying the fast order 2 FWER approximation. Also, there is potential to get quite close to the local significance level given by the full joint distribution of the test statistic vector by using order 3 or 4 approximation. Calculations using the more realistic empirical correlation matrix of part of chromosome 22 of the TOP study confirm this impression.

The maxT method (Section~\ref{maxT}) of \cite{WestfallYoung1993} may give higher power than FWER approximation (Table~\ref{tab:TOPres}). However, there is no general way of including environmental covariates using that method (Section~\ref{maxT}). Also, computing time is much larger than for lower-order FWER approximation (Section~\ref{sec:topvo2}), and the $\al$ estimate would likely differ if a new set of permutations were made (see confidence limits of Table~\ref{tab:TOPres}).

Another alternative is parametric bootstrap methods \citep{SeamanMyhsok2005}, which  could be used to estimate the local significance level when the exchangeability assumption is not satisfied. It would be an efficient method, but to our knowledge it has not been proven that parametric bootstrap will control the overall error rate, since nuisance parameters need to be estimated.

\cite{ConneelyBoehnke2007} introduced a method for multiple testing correction for GLMs for multiple responses (traits) based on the estimated correlation matrix of the score vector. The focus of the method is to calculate FWER-adjusted $p$-values based on the multivariate integral arising from \eqref{eq:FWER}. Currently, this integral can be computed numerically with some accuracy for dimensions smaller than or equal to 1000 using the \texttt{pmvnorm} function of the R package mvtnorm (see Section~\ref{sec:alphaksim} for details and references). Thus, the method of \citeauthor{ConneelyBoehnke2007} is not applicable for larger problems, e.g. more than $m=1000$ hypothesis tests. 

For our order $k$ FWER approximation method we have used standard R functions to compute the second order approximation given by~\eqref{eq:gamma2}. For orders 3, 4 and 5 we have used the above-mentioned function \texttt{pmvnorm} specifying the Miwa algorithm \citep{Miwa2003} instead of the default Genz--Bretz algorithm. The Miwa algorithm can be used for small dimensions, and is deterministic, whereas the Genz--Bretz algorithm includes simulations that lead to inaccuracies, which accumulate to an intolerable level when used for the large number of factors in~\eqref{gammak}. The research into better and faster integration of multivariate normal densities is ongoing, and \cite{Botev2016} provides an interesting new approach, applicable for dimensions smaller than or equal to 100. This will enable our order $k$ FWER approximation method to be applied with larger values of~$k$ than what has been presented here.

\section{Conclusions}\label{sec:conclude}
We have presented a new method for controlling the FWER for GWA data. The order~$k$ FWER approximation method can be used for generalized linear models and include adjustment for environmental covariates, possibly confounding, like population substructure or sex. We have applied the FWER approximation method to GWA data, and shown that our method is a powerful alternative to the Bonferroni and \Sidak methods, especially in situations were permutation methods cannot be used (exchangeability assumption not satisfied).

The method provides a local significance level, $\al$, for the individual tests, meaning that the null hypothesis of no association between phenotype and genetic marker should be rejected if the (unadjusted) $p$-value of a test is less than $\al$. We found a substantial increase in $\al$ already at the order~2 approximation, compared to the $\al$ produced by the well-known Bonferroni and the \v Sid\'ak methods -- methods that does not take correlation structure between the test statistics of the markers into account (\v Sid\'ak assumes independence, but that could be considered worst-case for GWA data).

\section*{Software}
The statistical analysis were performed using R \citep{R}, and the preprocessing of the genetic data were done using the software PLINK \citep{plink}.  

\section*{Acknowledgements}
The authors would like to thank Dr. Jelle J. Goeman (Leiden University Medical Centre, Leiden, The Netherlands) for valuable comments. Part of the work was done while the last author was on sabbatical at Centre for the Genetic Origins of Health and Disease, University of Western Australia, Australia. The PhD position of the first author is founded by the Liaison Committee between the Central Norway Regional Health Authority (RHA) and the Norwegian University of Science and Technology (NTNU).

\emph{Conflict of Interest:} None declared.

\bibliographystyle{chicago}
\bibliography{meff}

\end{document}